\documentclass{article}
\usepackage{spconf,amsmath,graphicx}
\usepackage{color,multirow}
\usepackage{cite}

\title{improving voice separation by incorporating\\
end-to-end speech recognition}
\name{
\begin{tabular}{c}
Naoya Takahashi$^{1,2}$, Mayank Kumar Singh$^{3*}$, Sakya Basak$^{4*}$, \\
Parthasaarathy Sudarsanam$^5$, Sriram Ganapathy$^4$, Yuki Mitsufuji$^1$
\end{tabular}
}

\address{$^1$Sony Corporation, Japan, \quad
         $^2$University of Tsukuba, Japan\\
         $^3$Indian Institute of Technology Bombay, India, \quad
         $^4$Indian Institute of Science, India\\
         $^5$Sony India Software Centre, India
         }
\begin{document}
\fontsize{9.5pt}{12pt}\selectfont
\maketitle
\begin{abstract}
Despite recent advances in voice separation methods, many challenges remain in realistic scenarios such as noisy recording and the limits of available data. In this work, we propose to explicitly incorporate the phonetic and linguistic nature of speech by taking a transfer learning approach using an end-to-end automatic speech recognition (E2EASR) system. The voice separation is conditioned on deep features extracted from E2EASR to cover the long-term dependence of phonetic aspects. Experimental results on speech separation and enhancement task on the AVSpeech dataset show that the proposed method significantly improves the signal-to-distortion ratio over the baseline model and even outperforms an audio visual model, that utilizes visual information of lip movements.
\end{abstract}
\section{Introduction}
\label{sec:intro}

\renewcommand{\thefootnote}{\fnsymbol{footnote}}
\footnote[0]{* this work was mainly performed when Singh and Basak were interns at Sony.}

The problem of separating a voice from other sources, such as speech separation (separating multiple overlapping speech signals)\cite{Hershey16,Kolbek17,Luo18cTAS,Takahashi19} and singing voice separation (separating vocals from other instrumental sounds)\cite{Takahashi18,Takahashi18MMDenseLSTM,Takahashi17} has been actively investigated for decades.
Various approaches including spectral clustering \cite{Bach06}, computational auditory scene analysis \cite{Hu13}, and non-negative matrix factorization (NMF)\cite{Schmidt06,VirtanenC09, MysoreS12, WangS14a} has been proposed to tackle this problem.
Recent advances in deep-learning-based methods have dramatically improved the accuracy of separation, and in some constrained scenarios, some methods perform nearly as well as or even better than ideal mask methods, which is used for theoretical upper baselines \cite{Luo18cTAS}. However, the performance drops significantly when we consider more challenging and realistic scenario such as separation in a noisy and reverberant environment \cite{Wichern19} or limited training data availability \cite{Takahashi18MMDenseLSTM}. 

There have been several attempts to improve voice separation in challenging scenarios by incorporating auxiliary information. In \cite{Afouras18,Ephrat18}, visual cues were used to incorporate a speaker's lip movement, which correlates to speech, and improvement over an audio only approach was observed in a noisy environment. However, such visual information is not always available owing to occlusion, low light levels or the unavailability of a camera.
In \cite{Kanda19,voice_activity}, guided (or informed) source separation was proposed where voice activity information was used to guide a voice separation model when the target speech is active. However, such binary information provides only marginal information and the phonetic aspect of speech is largely neglected.

In this work, we propose to explicitly incorporate the phonetic nature of a voice using a transfer learning approach. The transfer learning approach, where a deep neural network model is trained on a task in which a large dataset is available and hidden layer representation is transferred to another task, which might not have enough data, has been shown to be effective in an array of tasks including those in the audio domain \cite{Donahue2013,Tran2014,Takahashi2017AENet}. 
Here, we propose to utilize an end-to-end automatic speech recognition (E2EASR) system trained on a large speech corpus, and transfer deep features extracted from E2EASR (E2EASR features) to a voice separation task. E2EASR aims to combine all sub modules in an ASR system such as acoustic models, pronunciation models, and language models. Therefore, E2EASR is designed to model longer dependences, and hence E2EASR features are expected to contain longer linguistic information than conventional acoustic models which typically accept a few frames as input. If the phonetic and linguistic information is known, voice separation can more robustly estimate the target voice spectrogram under a noisy or unseen condition by leveraging the prior knowledge of phone-dependent spectrogram shape. For example, knowing that the current spectrum corresponds to fricatives, one can expect the target spectrum to contain high frequency components even when it is unclear from the noisy spectrogram.
There are prior works that used a conventional ASR for a speech enhancement task (separating speech from noise). Raj et.al. proposed phone-dependent NMF, where the bases of NMF were pre-trained on each phoneme and noise, and ASR was used to choose which bases to use for speech reconstruction \cite{Raj11}. In \cite{Wang16}, a DNN-based phoneme-specific voice separation approach was proposed, where the DNN models were trained for each phoneme and ASR was used to choose the model. However, these NMF bases and the DNN model selection approach treat each phoneme independently and largely ignore the context of the sequence of phonemes, making it difficult to incorporate long-term dependence. Moreover, such hard model switching mechanism largely relies on the accuracy of ASR and may produce artifacts due to misclassification or discontinuity. On the other hand, our method utilizes a single model to separate entire utterances and continuously incorporates phonetic information. This allows us to model longer dependences.

The contributions of this work are summarized as follows:
{
\setlength{\leftmargini}{20pt} 
\begin{enumerate}
\item We propose a transfer learning based approach to incorporate phonetic and linguistic nature of speech for voice separation. To this end, we propose the E2EASR features.  
\item We evaluated the proposed method on a simultaneous speech separation and enhancement task using AVSpeech and Audio-Set datasets, whose audio is recorded in non-controlled environments, and show that the proposed method significantly improves the separation accuracy over a model trained without E2EASR features and  a model trained with visual features.
\item We further show that even though E2EASR is trained on standard speech, it transferred robustly for a singing voice separation task with limited amount of data.
\end{enumerate}
}

\section{End-to-end ASR feature}
\label{sec:feature}
To capture phonetic and linguistic information, it is important to model the long-term dependences of an utterance. Conventional ASRs typically decompose the problem to acoustic modeling, pronunciation modeling, and linguistic modeling by assuming conditional independence between observations. The acoustic models typically accept only a few adjacent frames to estimate posterior probabilities of phonemes (or alternative subword units\cite{Takahashi16AAE}) for each frame, and then the hidden Markov model (HMM) is used to model the sequence of phonemes. However, this mechanism limits the modeling capabilities and requires expensive handcrafted linguistic resources such as a pronunciation dictionary, tokenization or phonetic context-dependency trees. E2EASR attempts to overcome these problems by combining all modules and learning unified model from only orthographic transcriptions which are easy to obtain. In \cite{hybrid_ctc_attention}, authors alleviated the conditional independence assumption by introducing a hybrid CTC/attention architecture, that allows the long term dependence to be modeled. This motivated us to use this model for transfer learning in voice separation task since we expect that the single model can provide phonetic and linguistic information. Furthermore, since the model is fully DNN-based, it has the potential to jointly train the voice separation and the E2EASR model by standard back-propagation. However, in this work, we adopt the deep feature approach since it is easy to use for many different tasks such as speech separation, speech enhancement, and singing voice separation.

\begin{figure}[t]
  \centering
  \includegraphics[width=\linewidth]{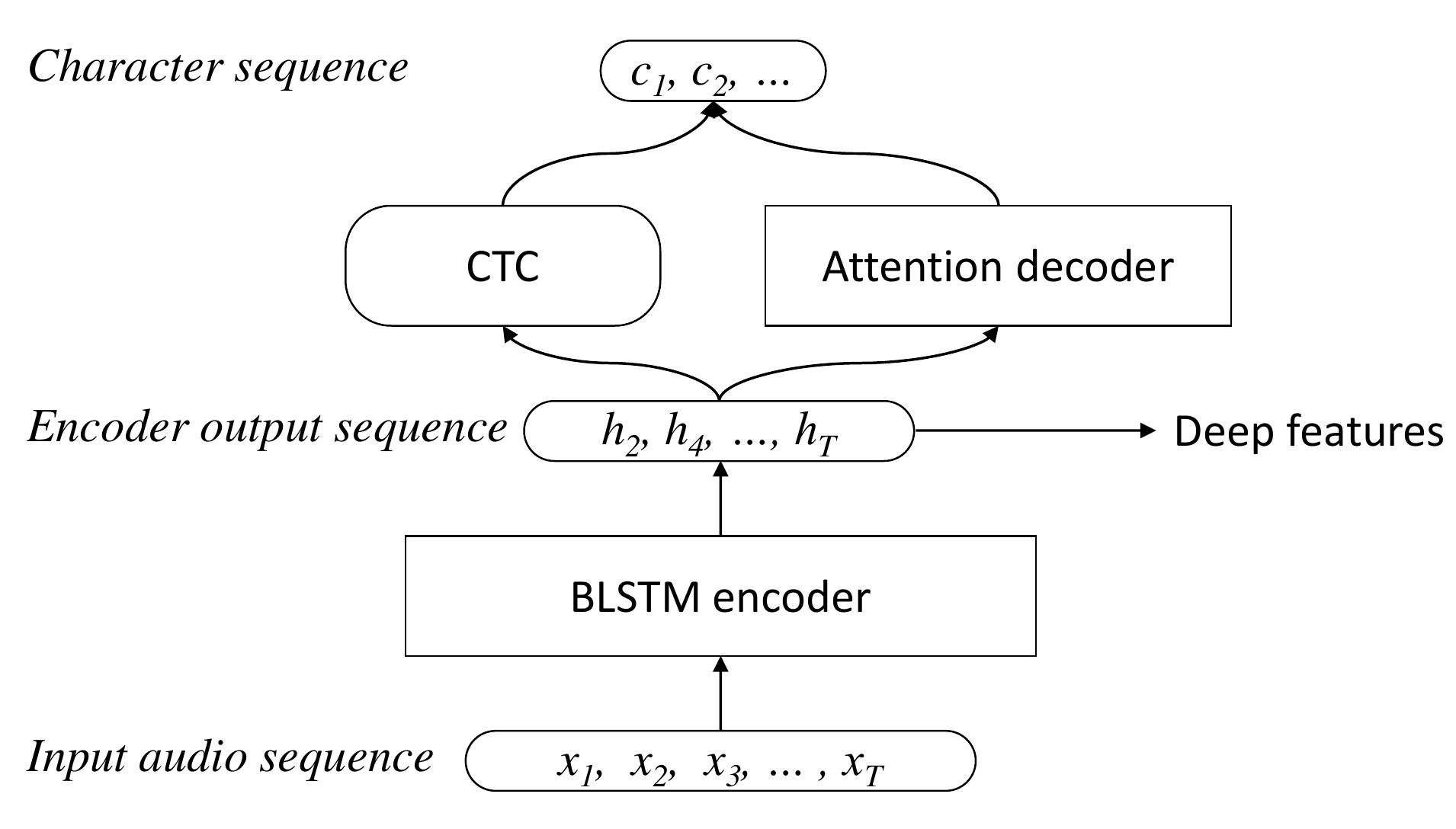}
  \caption{Illustration of E2EASR \cite{hybrid_ctc_attention}. The encoder output is used as deep features. Note that outputs of the CTC and attention decoder no longer have time alignment with the input. }
  \label{fig:asr}
\end{figure}

Deep features are a convenient yet powerful way of transfer learning for a DNN-based model. Many deep features in image \cite{Donahue2013}, video \cite{Tran2014} or audio domains \cite{Takahashi2017AENet}  are extracted from the activations of the last few fully connected (fc) layers of DNN models trained on a classification task in which the input size is fixed. Here, we want the E2EASR features to maintain its time resolution while containing long term dependency. Therefore, deep features are extracted from the encoder output to preserve the time alignment, as shown in Figure \ref{fig:asr}. Another possibility could be to use attention weights as deep features, which we leave as a future work.

\section{Voice Separation with E2EASR feature}
\label{sec:SS}
\begin{figure}[t]
  \centering
  \includegraphics[width=\linewidth]{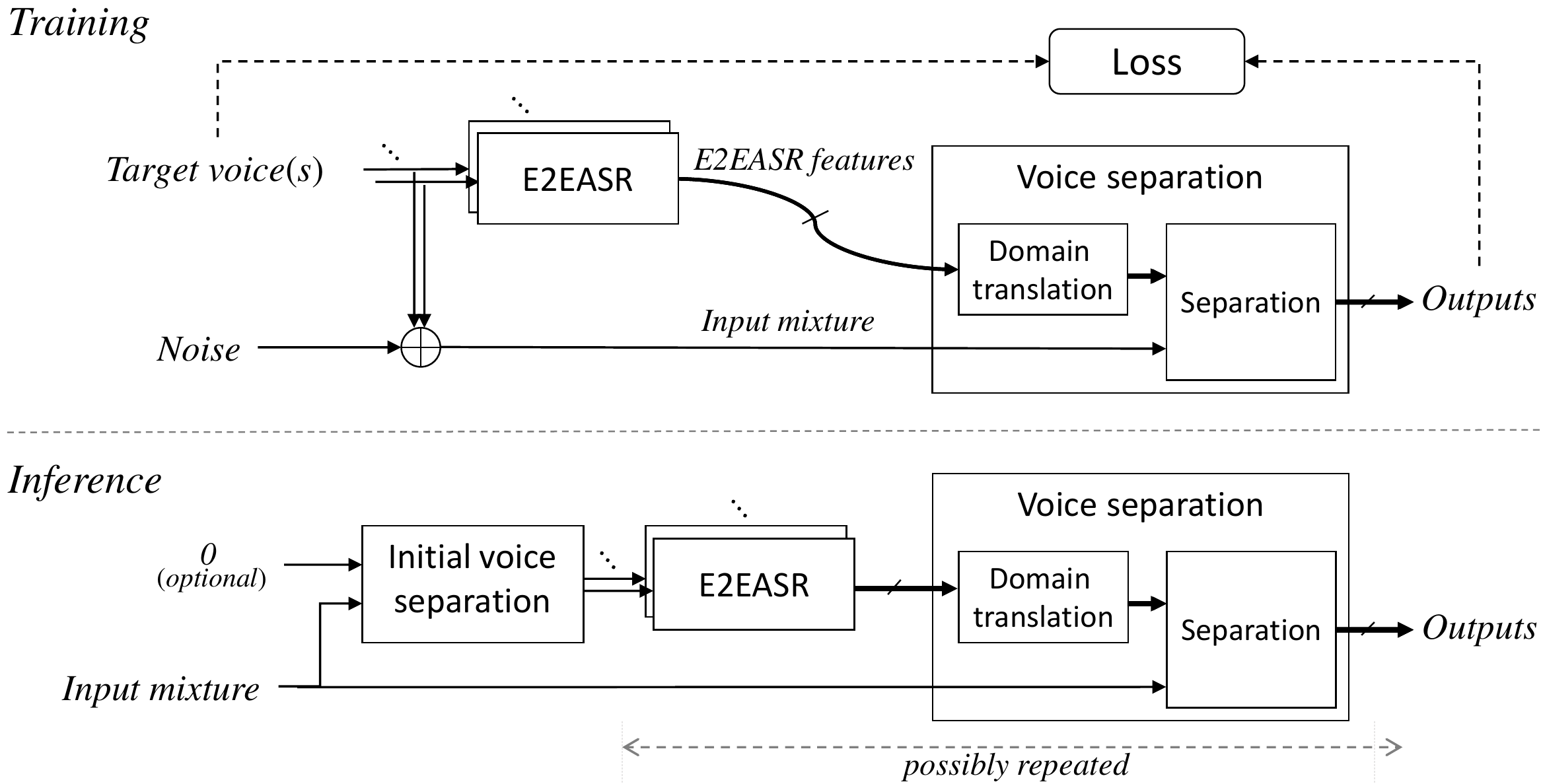}
  \caption{Illustration of proposed method. E2EASR features are extracted from all voice signals. 
}
  \label{fig:vs}
\end{figure}

The voice separation process using E2EASR features is illustrated in Figure \ref{fig:vs}.
The proposed E2EASR features are passed to the voice separation model along with the input mixture to incorporate phonetic and linguistic information. The source separation model is equipped with a domain translation network to convert the E2EASR feature to a suitable representation for voice separation, and the output of the domain translation network is concatenated with audio along the channel dimension. The domain translation network is trained simultaneously for the voice separation task. During the training, we extracted the E2EASR features from the target (oracle) speech. During the inference time, since the target speech is unknown, we first separate the voice without using E2EASR features. 
For this, we can use another voice separation model that does not use E2EASR features. However, to avoid the need for an additional model, we can instead use the model that uses E2EASR features but feed zero data as E2EASR features for the initial stage. 
Since deep features tend to have sparse representation, feeding zero data as E2EASR features does not disrupt results. We found that this approach provides comparable results with the model trained with audio only. After the initial voice separation, we extract the E2EASR features from the separated voices and feed the feature to the voice separation model again with the input mixture.
Note that we used the oracle target speech during the training time to avoid the voice separation model recursively depending on its separation quality through the E2EASR features from separated sources. This could cause a mismatch in the E2EASR feature quality between the training and inference times since the estimated source is used for inference. To close the gap, we can iterate the voice separation and E2EASR feature extraction to progressively enhance the feature quality.

\section{Experiments}
We evaluated the proposed approach on a speech separation task and singing voice separation (SVS) task. The experiments on speech separation was designed to evaluate the noisy speech scenario while the experiments on SVS examined the case of limited data availability, which is really the case in the community. 

\subsection{Speech separation}


\label{sec:exSS}
\subsubsection{Datasets}


To evaluate the proposed method under a realistic and challenging environment, we first conducted experiments on a single-channel speaker-independent simultaneous speech separation and enhancement task, where the goal is to separate multiple overlapping speech while removing the noise. For the speech dataset, we used the AVSpeech dataset \cite{Ephrat18}, which consists of 4700 hours of YouTube videos of lectures and how-to videos. We used a subset of the dataset: 100 hours for training and 15 hours for testing. The audio in the AVSpeech dataset was recorded in less-controlled environments than that in the WSJ0 dataset, which is often used to evaluate speech separation methods, making the task more challenging. Indeed, we found that a speech separation model that performs well on the WSJ0 dataset performs poorly on the AVSpeech dataset. To make the problem more challenging, we also added noise from the AudioSet dataset\cite{AudioSet}, which is a collection of 10-second sound clips drawn from YouTube videos. We omitted the classes likely to contain human voices. After processing, the training and testing sets consisted of 105 and 11 hours of audio, respectively.
The preprocessing of data was done similar to \cite{Ephrat18}, where 3 second, sequential, non-overlapping crops of audio from the AVSpeech and Audio-Set datasets were extracted and normalized to have a maximum amplitude of one for the speech and 0.3 for the noise, respectively. Mixtures were created by randomly mixing two different speakers and the noise.
The LibriSpeech corpus\cite{librispeech}, which consists of 960 hours of speech, was used for training E2EASR.



\subsubsection{Model Architecture}
\begin{figure}[t]
  \centering
  \includegraphics[width=80mm]{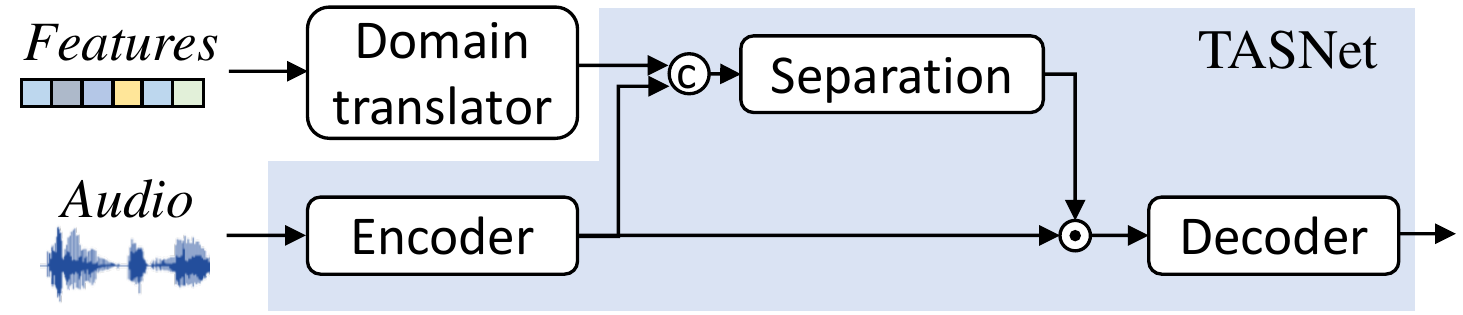}
  \caption{TASNet based voice separation network}
  \label{fig:arch}
\end{figure}

For the voice separation network architecture, we adopted TASNet \cite{Luo18cTAS}, which is a recently proposed time domain speech separation network that produced state-of-the-art results on the WSJ-2mix and WSJ-3mix datasets. Architecture details are illustrated in Figure \ref{fig:arch}. E2EASR features were interpolated after passing through the domain translation network to have the same frame rate as the encoder outputs of TASNet. The domain translation network comprised six 1-D convolution layers with $256$ filters and a filter size of $5$.  Then, its output was concatenated with the encoder output and passed to the separation branch for mask prediction.
The architectures of encoder, decoder and separation networks in TASNet were same as the \textit{Conv-TasNet-gLN} in \cite{Luo18cTAS} except the $B$ (number of channels in bottleneck 1$\times$1-conv block), which was increased 512 to provide extra capacity to incorporate the E2EASR features.
The network was trained with the permutation invariant training (PIT) with SI-SDR criteria as in \cite{Luo18cTAS}.
 We used the baseline visual feature model, which is described in the next section, as the initial voice separation model.
For E2EASR, we used ESPnet framework \cite{espnet}, in which the hybrid CTC/attention model \cite{hybrid_ctc_attention} is available. The dimension of the E2EASR feature was 1024.


\subsubsection{Baselines}
We considered two baseline models. For the lower baseline, we trained TASNet without extra features using permutation invariant training (PIT). To assess the quality of the E2EASR features, we also considered visual features extracted as follows: the lip regions were cropped from video after the faces were aligned using the facial landmarks derived in \cite{lipcrop} and reshaped to dimensions of (96, 96). 
Then an autoencoder with an architecture of 3 convolution layers followed by 2 linear layers and 3 transposed convolution layers was trained on lip images at the frame level. 
The bottleneck layer activations were used as visual deep features. We also tried using the visual features in \cite{Ephrat18,Afouras18}, however we found that these were less effective due to the unavailability of the training data used in \cite{Ephrat18,Afouras18}. 
The network architectures
were the same as the one in the proposed model with E2EASR features.

\subsubsection{Results}
Table \ref{tab:se} compares the scale-invariant signal-to-distortion ratio improvement (SI-SDRi) \cite{Luo18cTAS}. As shown in the table, providing additional features to the source separation model generally improves the performance. The visual feature provides $1.2$ dB improvement over the audio only lower baseline model. This result is similar to \cite{Ephrat18}, although the metric is slightly different. Our proposed method significantly outperformed the visual feature model, providing 3.5 dB improvement over the lower baseline. This is somewhat surprising since the proposed method significantly outperformed the audio visual model with a single modality.  
To further assess the effects of the E2EASR features quality mismatch between the training and inference, we also evaluated the oracle E2EASR features, where the features were extracted from the target signal, which is not available in a real scenario. The SI-SDRi difference between the oracle and estimated E2EASR features was only 0.2 dB. This indicates that E2EASR features are robustly extracted from degraded voices, and are therefore less sensitive to the quality of the initial voice separation model.

\begin{table}[t]
    \caption{\label{tab:se} {\it Comparison of SI-SDR improvement for speech separation.}}
    \vspace{2mm}
    \centering{
    \begin{tabular}{ c |c } 
    \hline
    Method	&  SI-SDRi[dB] \\
    \hline\hline
    No extra features (PIT) &  9.0  \\ 
    Visual features &  10.2  \\ 
    Estimated E2EASR feature     &  \textbf{12.5}  \\ 
    \hline
    Oracle E2EASR feature      & 12.7  \\ 
    \hline
    \end{tabular}
    }
\end{table}

\subsection{Singing Voice Separation}

\label{sec:exSVS}
\subsubsection{Datasets}
We further evaluated the proposed method on an SVS task using MUSDB dataset, prepared for SiSEC 2018 \cite{sisec2018}. MUSDB has 100 and 50 songs each in the {\it Dev} and {\it Test} sets, respectively recorded in stereo format at a 44.1kHz sampling rate. Even though MUSDB is one of the largest professionally recorded dataset for the SVS task, it has only about 6.7 hours of data for training, which is very small compared with the AVspeech dataset, which has about 4700 hours of data. 
Moreover, this task was much more difficult since the E2EASR model was trained on standard speech then transferred to singing voice data which has significant domain mismatch as it is more variable in the fundamental frequency range, timbre, tempo, and dynamics.

\subsubsection{Setup}
We used multi-scale DenseNet (MDenseNet) \cite{Takahashi17} for the voice separation network. The number of layers $l$ and growth rate $k$ of each scale in MDenseNet are described in Table \ref{tab:mdense}. Since our goal is not to achieve state-of-the-art performance but to show the effectiveness of the E2EASR features, we converted the input magnitude spectrogram to a 128-band mel spectrogram to reduce the dimension. The E2EASR features were passed to the domain translation network and concatenated to the audio after the initial convolution layer of MDenseNet. The network was trained to minimize mean square error in the mel spectrogram domain. During the inference time, the output mel spectrogram was converted to wave domain by using a pseudo-inverse and the input phase. 
For the initial voice separation, we used the same model by providing all zero values as dummy E2EASR features. This allows us to avoid the need for an additional model. 
We used the same E2EASR model as in the previous experiment and it was not trained on a singing voice dataset.

\begin{table}[t]
\footnotesize
\caption{\label{tab:mdense} {\it Separation network based on MDenseNet\cite{Takahashi17}.}}
\centering{
\begin{tabular}{c c c c c c c c} 
\hline
scale & 1 & $\frac{1}{2}$ & $\frac{1}{4}$ & $\frac{1}{8}$ & $\frac{1}{4}$ & $\frac{1}{2}$  & 1\\
\hline
$l$	&  4 &	 4	&  4 &	4  &  4	&  4 & 4\\   
$k$	& 14 &	15	& 16  & 17 & 16	& 15 & 14 \\   
\hline
\end{tabular}
}
\end{table}

\subsubsection{Results}
We calculated the SDR improvement using the {\it museval} package \cite{sisec2018} and compared in Table \ref{tab:svs}. 
As baselines, we consider the MDenseNet model trained without E2EASR features. 
Even though E2EASR was not trained on singing voice data during the training, the E2EASR features improved SDR over the baseline. Moreover, the difference between the oracle and estimated E2EASR features is marginal, indicating the robustness of E2EASR features against the initial voice separation quality.

\begin{table}[t]
    \caption{\label{tab:svs} {\it Comparison of SDR improvement for SVS.}}
    \vspace{2mm}
    \centering{
    \begin{tabular}{ c | c} 
    \hline
    Method	&  SDRi[dB] \\
    \hline\hline
    Baseline (w/o feature) &   8.65 \\ 
    Estimated E2EASR feature &  8.80 \\ 
    Oracle E2EASR feature    &  8.82 \\ 
    \hline
    \end{tabular}
    }
\end{table} 


\section{Conclusion}
We proposed a transfer learning approach to leverage the E2EASR model for voice separation. Experimental results on the simultaneous speech separation and enhancement task shows that the proposed E2EASR features provide a significant improvement over baselines including  a model using visual cues. We further show that the E2EASR features improve the performance of SVS, demonstrating its robustness against limited data availability and domain mismatch.

\ninept

\bibliographystyle{IEEEbib}
\bibliography{bss,other}

\end{document}